\documentclass{article}

\usepackage{PRIMEarxiv}
\usepackage{makecell}
\usepackage[utf8]{inputenc} 
\usepackage[T1]{fontenc}    
\usepackage{hyperref}       
\usepackage{url}            
\usepackage{booktabs}       
\usepackage{amsfonts}       
\usepackage{nicefrac}       
\usepackage{microtype}      
\usepackage{lipsum}
\usepackage{multirow}
\usepackage{graphicx}
\graphicspath{{media/}}     
\usepackage{longtable}
\usepackage[flushleft]{threeparttable}
  \usepackage{xcolor,colortbl}
\title{Towards Automated Quality Assurance of Patent Specifications: A Multi-Dimensional LLM Framework}

\author{
  Yuqian Chai, Chaochao Wang, Weilei Wang \\
  Patsnap \\
  Suzhou\\
  \texttt{\{chaiyuqian, wangchaochao, wangweilei\}@patsnap.com} \\
}

\begin{document}
\maketitle

\begin{abstract}

Although AI drafting tools have gained prominence in patent writing, the systematic evaluation of AI-generated patent content quality represents a significant research gap. To address this gap, We propose to evaluate patents using regulatory compliance, technical coherence, and figure-reference consistency detection modules, and then generate improvement suggestions via an integration module. The framework is validated on a comprehensive dataset comprising 80 human-authored and 80 AI-generated patents from two patent drafting tools. Evaluation is performed on 10,841 total sentences, 8,924 non-template sentences, and 554 patent figures for the three detection modules respectively, achieving balanced accuracies of 99.74\%, 82.12\%, and 91.2\% against expert annotations. Additional analysis was conducted to examine defect distributions across patent sections, technical domains, and authoring sources. Section-based analysis indicates that figure-text consistency and technical detail precision require particular attention. Mechanical Engineering and Construction show more claim-specification inconsistencies due to complex technical documentation requirements. 
AI-generated patents show a significant gap compared to human-authored ones. While human-authored patents primarily contain surface-level errors like typos, AI-generated patents exhibit more structural defects in figure-text alignment and cross-references.

\end{abstract}

\section{Introduction}
\label{sec:intro}

With over 3.55 million patent applications filed globally in 2023\footnote{\url{https://www.wipo.int/pressroom/en/articles/2024/article_0015.html}, accessed August 2025}, the patent system faces an unprecedented scale challenge. While expert review remains the gold standard for quality assessment, scalability, cost, and consistency issues make it increasingly unsustainable. Meanwhile, the rise of AI-powered patent drafting tools has further amplified the need for automated quality assessment solutions \cite{rehnstrom2024drafting,mehrotra2024navigating}.

Patent specifications serve as the legal foundation that defines invention scope and boundaries \cite{bui2025advancing, khera2025can}. High-quality specifications must meet stringent requirements including regulatory compliance, technical accuracy, logical consistency, and proper figure integration. Quality defects can result in patent rejection, costly prosecution delays, or inadequate protection that undermines IP value. Traditional quality assurance relies on manual review, which is thorough but increasingly unsustainable given growing application volumes and cost constraints.

Meanwhile, the widespread adoption of AI-powered drafting tools has paradoxically intensified this challenge. Despite enhancing drafting efficiency, they introduce new quality risks including technical inaccuracies, inadequate disclosure, and improper legal terminology \cite{mehrotra2024navigating,ray2023ai}. Such defects not only threaten legal validity but also undermine patent disclosure's role in advancing technological knowledge \cite{wang2024patentformer, ren2025large}. These AI-specific quality risks, combined with scalability pressures, create an urgent need for automated quality assessment solutions.

However, current patent quality assessment methods fall short of addressing these challenges. Traditional methods rely heavily on quantitative proxy indicators such as citation patterns and legal status \cite{burke2007measuring, harhoff2003citations}, while recent AI-driven approaches employ machine learning models to predict patent value \cite{hu2023evaluation, trappey2019patent}. These existing methods suffer from performance degradation across different technical fields, and neglect document-intrinsic quality \cite{zhang2024research}. 
While some recent work has begun to consider document content, these studies fail to address patent-specific requirements, such as legal compliance and technical accuracy. For instance, Wang et al. \cite{wang2024autopatent} evaluate specification quality using generic metrics like ROUGE and BLEU. This limitation highlights the need for an evaluation framework tailored specifically for patent specifications.

While LLMs have demonstrated remarkable capabilities in patent writing \cite{wang2024autopatent}, their potential for patent quality assessment remains largely unexplored. With cross-domain knowledge and reasoning capabilities, LLMs can both analyze patent content for various issues and provide actionable feedback. Recent advances in multimodal LLMs further enhance this potential by enabling integrated analysis of both textual and visual elements in patent specifications \cite{zhao2023survey}. This potential has been explored in recent studies, including work by Yoo et al. \cite{yoo2025patentscore} who leveraged LLMs to assess AI-generated claims. However, existing research has primarily focused on patent claims rather than specifications, leaving a significant gap in comprehensive patent quality evaluation.

This paper introduces the first comprehensive automated quality assurance framework specifically designed for patent specifications. Our key contributions are as follows: 

(1) \textbf{Three-dimensional assessment framework}: An LLM-driven quality evaluation system that assesses regulatory compliance, technical content coherence, and figure-text consistency;

(2) \textbf{Actionable improvement system}: Beyond defect detection, our framework provides specific, actionable recommendations for specification improvement, enabling iterative quality enhancement;

(3) \textbf{Large-scale empirical validation}: Comprehensive evaluation on 160 patent specifications across eight technical domains, demonstrating robust performance across diverse patent types and complexity levels;

(4) \textbf{Defect pattern identification}: distinct quality issues across patent sections, technical domains, and sources.


The remainder of this paper is organized as follows: Section \ref{sec:related} reviews related work in AI-generated content detection and patent quality assessment. Section \ref{sec:method} details our methodology and framework design. Section \ref{sec:setup} describes our experimental setup, including dataset construction and evaluation metrics. Section \ref{sec:results} presents experimental results and analyzes defect patterns across patent sections, technical domains, and authoring sources. Finally, Section \ref{sec:conclusion} concludes the paper and summarizes our contributions.

\section{Related Work}
\label{sec:related}

\subsection{LLM Applications in Patent Writing}

Research and commercial platforms have explored LLMs' capabilities in patent writing from component-level generation to full document automation. 
For instance, studies have investigated claim and abstract generation tasks using GPT-2 and Llama-3 \cite{jiang2024can, lee2020patent}.
 Additionally, automated claim revision systems have been developed to assist in the transition between application and granted versions \cite{jiang2024patent}. Multi-agent frameworks like AutoPatent \cite{wang2024autopatent} distribute patent writing tasks across specialized agents, showing improved performance over single-model approaches. In contrast to these component-level academic approaches, commercial platforms have achieved full-document generation capabilities, including MindFlowing\footnote{\url{https://www.mindflowing.cn/}, accessed August 2025} and PatSnap\footnote{\url{https://www.patsnap.com/}, accessed August 2025}. However, comprehensive evaluations for accuracy and reliability remain notably absent from the literature. This evaluation gap is concerning given the critical nature of patent documentation, where defects can have significant legal and financial consequences.

\subsection{Automated Patent Assessment}

Human-based patent quality evaluation is plagued by high costs, lengthy processes, and subjective judgments, thereby driving the need for automated assessment methods \cite{burke2007measuring}. Early automated approaches relied on quantitative proxy indicators, including citation patterns, patent family characteristics, claims scope, and legal status \cite{burke2007measuring, harhoff2003citations}. Later research developed more comprehensive frameworks incorporating technical, legal, and economic attributes \cite{hu2023evaluation, zhang2024research,grimaldi2020indexes}. For instance, Cricelli et al. \cite{cricelli2021patent} systematically identified 41 key performance indicators to assess patent. More recently, studies have investigated patent value prediction using machine learning and deep learning approaches, demonstrating promising results \cite{hu2023evaluation,wu2016patent,trappey2019patent}.
Furthermore, Wang et al. \cite{wang2024autopatent} used general text similarity metrics (BLEU, ROUGE) for LLM-generated patents, while Jiang et al. \cite{jiang2025towards} and Yoo et al. \cite{yoo2025patentscore} utilized LLMs to evaluate claim quality. Other approaches have explored semantic analysis for patent similarity assessment \cite{gerken2012new} and deep learning methods for explainable novelty detection \cite{jang2023explainable}. However, existing methods are constrained by poor cross-domain generalization, limited improvement actionability, and the absence of comprehensive evaluation frameworks for patent specification assessment \cite{zhang2024research,du2021personalized}.

\subsection{AI-Generated Content Detection}

The rapid proliferation of AI-powered writing tools has raised concerns about content authenticity and quality assurance, leading to the emergence of AI-generated content (AIGC) detection research \cite{wu2024surveyllmgeneratedtextdetection}. Current AIGC detection methods can be broadly categorized into three approaches: statistical methods \cite{arase2013machine}, neural classifiers \cite{wu2024detectrl, mitchell2023detectgpt}, and watermarking techniques \cite{kirchenbauer2023watermark}. Statistical methods analyze token probability distributions and syntactic complexity patterns to identify AI-generated content \cite{arase2013machine}. Neural detection approaches train binary classifiers using large corpora of human and AI-generated texts \cite{wu2024detectrl, mitchell2023detectgpt}. Watermarking techniques embed imperceptible signals during the generation process to enable later identification \cite{kirchenbauer2023watermark}. However, these methods face significant challenges in real-world deployment. Specifically, performance often degrades substantially when applied outside controlled experimental settings. Furthermore, binary classification approaches cannot provide the detailed analysis required for comprehensive patent evaluation. These limitations suggest a need for alternative approaches to patent quality assessment.

Patent evaluation systems should focus on constructive quality assessment by identifying specific deficiencies and providing improvement guidance. This shift from detection to assistance aligns better with the collaborative nature of patent development by focusing on enhancing specification quality regardless of authorship.

\section{Methodology}
\label{sec:method}

\begin{figure}
    \centering
    \includegraphics[width=1\linewidth]{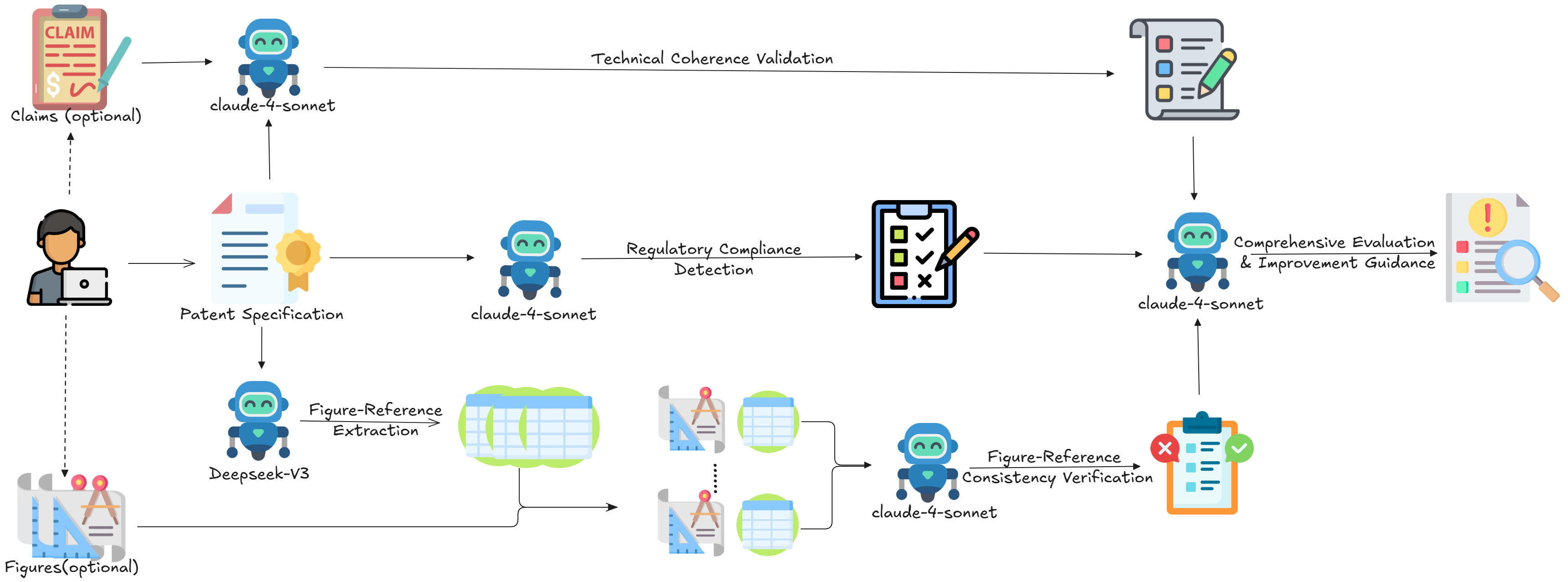}
    \caption{Overview of the patent quality assurance framework. Three specialized detection modules operate in parallel to identify regulatory compliance, technical coherence, and figure-reference consistency issues, followed by a sequential integration module that synthesizes findings into comprehensive improvement recommendations.}
    \label{fig:aigc}
\end{figure}

As displayed in Figure \ref{fig:aigc}, our framework consists of three parallel specialized detection modules followed by an integration module. The system accepts flexible input combinations from patent applications, including any available patent specification sections, patent claims, and figures.

 \subsection{Regulatory Compliance Detection}

Patent applications face increasingly complex regulatory requirements that vary substantially across jurisdictions, making compliance a critical yet challenging aspect. To systematically address these challenges, our module identifies six primary categories of regulatory violations commonly encountered in patent specifications.
These categories address both content and format requirements: (1) prohibited commercial language, such as subjective claims of "revolutionary breakthrough"; (2) inconsistent technical terminology and naming conventions; (3) missing mandatory sections required by specific jurisdictions; (4) improper title and abstract formatting; (5) orthographic errors that may affect legal interpretation; and (6) insufficient figure descriptions that compromise disclosure adequacy.

For each sentence, the module provides binary compliance assessments accompanied by jurisdiction-specific explanations. The module's adaptive framework dynamically adjusts detection criteria based on target patent office requirements, ensuring robust regulatory compliance across diverse jurisdictional contexts.

 \subsection{Technical Coherence Validation}
 
Assessing technical coherence in patent applications has become increasingly complex, requiring evaluation of both traditional quality issues and AI-generated defects. Conventional coherence problems include contradictions between claims and specifications, insufficient technical support, terminology inconsistencies, logical gaps that affect reproducibility, and conflicts with prior art. AI-generated patents compound these challenges by introducing fabricated technical details, artificially constructed reasoning sequences, and superficial domain descriptions that lack genuine depth.
The technical coherence validation module specifically targets the aforementioned issues.

The module evaluates each sentence using a four-tier risk classification (Safe to High Risk), with cross-validation to ensure reliability. High-risk violations, encompassing both traditional technical contradictions and AI-generated fabrications that receive priority attention for manual review, such as technical field mismatching. Medium-risk issues include insufficient support for claims, incomplete technical features, and vague parameters. Low-risk cases involve minor expression differences, typos, and technically equivalent expressions that typically need minimal intervention. Detailed classification examples are presented in Table~\ref{tab:risk_classification}.

\begin{table}[h]
\caption{Risk Classification Criteria and Examples for Technical Coherence Validation}
\label{tab:risk_classification}
\centering
\begin{tabular}{p{5.2cm}|p{5.2cm}|p{5.2cm}}
\hline
\textbf{High Risk} & \textbf{Medium Risk} & \textbf{Low Risk} \\
\hline
 • Contradicts fundamental theories
\newline • Includes fabricated technical details
\newline • Mismatches Technical field
\newline • Lacks technical details in claims
\newline • Lacks essential figures
\newline • Shows completely inconsistent drawing labels or descriptions
\newline • Constructs artificial reasoning chains
\newline • ....
& • Defines field scope imprecisely
\newline • Cites literature with low relevance
\newline • Provides insufficient support for claims
\newline • Lacks depth in technical descriptions
\newline • Establishes improper logical connections
\newline • Lacks proper experimental validation
\newline • Omits necessary views or includes irrelevant ones
\newline • ....
&  • Cites dated but valuable literature
\newline • Shows minor expression differences from claims
\newline • Contains typos not affecting understanding 
\newline • Has minor formatting issues
\newline • Uses slightly imprecise terminology
\newline • ....
\\
\hline
\end{tabular}
\end{table}

 \subsection{Figure-Reference Consistency Verification}

Figure-text misalignment represents a critical yet systematically under-addressed patent quality risk. Unlike the sentence-level processing of previous modules, the current module operates at the figure-reference pair level. Figure-reference pairs are classified as consistent when descriptions accurately reflect figure content, reference numbers correctly correspond to elements, and no extraneous references are introduced. This definition identifies common figure-related prosecution issues, including missing reference numerals, mismatched descriptions, and non-existent figure references.

The figure-reference consistency module performs three tasks to address the complex interdependencies inherent in patent figure sets. The first task identifies figure-related descriptions and constructs comprehensive reference numeral inventories through specification analysis. The second task verifies each figure against its corresponding textual description.  The last task implements cross-figure validation to prevent misidentification of legitimate variations as missing numerals. This pipeline ensures comprehensive detection of all figure-related textual elements with high accuracy.

 \subsection{Comprehensive Evaluation \& Improvement Guidance}

This module represents the system's final analytical stage, integrating outputs from all detection modules into unified filing readiness determinations. Each input contains categorized issues with severity that enable systematic readiness assessment.
The evaluation report employs a proprietary four-level readiness classification (Filing Ready, Minor Revision Needed, Major Revision Needed, Significant Rework Required) based on prosecution impact probability. Beyond classification, the module delivers comprehensive actionable intelligence with root cause analysis and modification recommendations, systematically organized by chapter sequence and severity level. This module bridges the gap between problem identification and practical resolution.

\section{Experimental Setup} \label{sec:setup}

\subsection{Research Design Overview}

This study assesses the effectiveness of automated patent quality evaluation using a balanced dataset of 160 patent documents. The dataset comprises human-authored and AI-generated patents spanning eight technical domains (detailed in Section \ref{sec:dataset}). We apply our three-module detection system to assess patent quality and validate results against expert annotations for ground truth establishment. 

The experimental design addresses the following research questions: 

RQ1: What is the detection accuracy of our framework compared to expert annotations?

RQ2: How do defect patterns vary across patent sections and technical domains?

RQ3: What are the key quality differences between human-authored and AI-generated patents?





\subsection{Dataset Construction Strategy}\label{sec:dataset}

\subsubsection{Sample Selection Criteria}

Human-authored patents represent established professional standards and serve as comparative baselines in this study, while we acknowledge that they may exhibit inherent defects. Three major criteria were applied for selecting human-authored patents. 
First, publication dates must fall between 2015 and October 2022 to avoid potential ChatGPT influence and establish pre-AI professional baselines. Second, patents must include complete specifications with clear technical disclosure and comprehensive claim sets. 
Third, patents were sampled to ensure balanced representation across eight technical domains and patent types.

We generated 80 AI patent specifications using a matched-pair design, where each AI document corresponds to a human-authored patent in Table \ref{tab:human_patent_distribution}. This design controls for confounding variables including technical complexity, domain expertise requirements, and solution sophistication. 
We generated documents evenly using Eureka and MindFlowing patent drafting tools, with 5 documents per IPC category and 20 documents per patent type. While each tool employs different internal prompting mechanisms and generation algorithms, we provided identical input materials (abstract, claims, technical descriptions, and figures) for each patent pair. 
Eureka generates complete specifications after receiving all input materials, while MindFlowing generates the major specifications  without figures. Additionally, the Brief Description of Drawings section generated by MindFlowing contains unusable figure codes. We therefore extracted the relevant figure descriptions from the narrative content to manually complete this section.

\begin{table}[htbp]
\centering
\caption{Human Patent Distribution by IPC Category and Patent Type}
\label{tab:human_patent_distribution}
\begin{tabular}{lccc}
\hline
\textbf{IPC Category} & \textbf{Invention Patents} & \textbf{Utility Model Patents} & \textbf{Total} \\
\hline
A - Human Necessities & 5 & 5 & 10 \\
B - Operations, Transport & 5 & 5 & 10 \\
C - Chemistry, Metallurgy & 5 & 5 & 10 \\
D - Textiles, Paper & 5 & 5 & 10 \\
E - Fixed Constructions & 5 & 5 & 10 \\
F - Mechanical Engineering & 5 & 5 & 10 \\
G - Physics & 5 & 5 & 10 \\
H - Electricity & 5 & 5 & 10 \\
\hline
\textbf{Total} & \textbf{40} & \textbf{40} & \textbf{80} \\
\hline
\end{tabular}
\end{table}

\subsubsection{Dataset Characteristics}

Table~\ref{tab:descriptive_stats} presents the descriptive statistics of our dataset. Document length shows the most significant variation across sources. Human-authored patents exhibit substantial variability (mean: 5,330 words, SD: 3,670), with a right-skewed distribution indicated by the mean-median difference. Eureka produces documents with length characteristics similar to human patents. In contrast, MindFlowing generates consistently shorter documents with minimal variation (mean: 2,756 words, SD: 586). This standardization indicates that MindFlowing has limited adaptability to the complexity and scope of the input technical materials. Claims count varies modestly across sources, ranging from 7.47 (Eureka) to 8.28 (MindFlowing), with human patents at 7.88. Figure references remain relatively consistent across all sources, with means ranging from 3.20 to 3.73. 
The paired design (80 human-AI patent pairs) ensures that observed differences reflect authorship capabilities rather than content variations.

\begin{table}[htbp]
\centering
\caption{Descriptive Statistics by Document Source}
\label{tab:descriptive_stats}
\begin{tabular}{lccc}
\hline
\textbf{Metric} & \textbf{Human Patents} & \textbf{Eureka} & \textbf{MindFlowing} \\
\hline
\multicolumn{4}{l}{\textit{Document Length (words)}} \\
Mean & 5329.88 & 4577.35 & 2756.38  \\
Median &3670.00 & 3907.00 & 2585.50   \\
SD & 3670.00 & 3907.00 & 2585.50  \\
\hline
\multicolumn{4}{l}{\textit{Claims Count}} \\
Mean & 7.88 & 7.47 & 8.28 \\
Median & 8.00 & 7.00 & 9.50 \\
SD & 3.22 & 3.76 & 2.56 \\
\hline
\multicolumn{4}{l}{\textit{Figure References}} \\
Mean & 3.46 & 3.20 & 3.73 \\
Median & 3.00 & 3.00 & 3.00 \\
SD & 2.39 & 2.81 & 1.88 \\
\hline
\multicolumn{4}{l}{\textit{Sample Size}} \\
N & 80 & 40 & 40 \\
\hline
\end{tabular}
\end{table}

\subsection{Evaluation Methodology}

To validate our patent quality assessment framework, we evaluate the three detection modules using balanced accuracy, precision, recall, and F1-score metrics. Specifically, balanced accuracy is employed because all detection modules exhibit varying degrees of class imbalance, such as imbalanced compliance ratios and uneven risk level distributions. Balanced accuracy provides equal weight to each class's performance, thereby offering a more reliable assessment across all categories. All patents were systematically anonymized to ensure objective assessment during the expert annotation and framework evaluation phases, while maintaining technical content integrity.

\section{Results and Analysis}
\label{sec:results}
\subsection{Overall Framework Performance (RQ1)}
Table~\ref{tab:detection_summary} presents the comprehensive performance of our three detection modules. We report balanced accuracy instead of standard accuracy to address the significant class imbalance in our datasets, where minority classes represent only 5-33\% of samples. 

\begin{table}[htbp]
\centering
\caption{Framework Detection Performance Summary}
\label{tab:detection_summary}
\begin{tabular}{lccccc}
\toprule
\textbf{Task \& Category} & \textbf{N} & \textbf{Balanced Acc.} & \textbf{Precision} & \textbf{Recall} & \textbf{F1-Score} \\
 & & \textbf{(\%)} & \textbf{(\%)} & \textbf{(\%)} & \textbf{(\%)} \\
\midrule
\textbf{Regulatory Compliance} & 10,841 & 99.74±0.07 & & & \\
\quad Compliant & 10,298 & & 100.0±0.00 & 99.49±0.14 & 99.74±0.07 \\
\quad Non-compliant & 543 & & 91.17±2.11 & 100.0±0.00 & 95.19±1.20 \\
\addlinespace
\textbf{Technical Coherence} & 8,924 & 82.12±2.28 & & & \\
\quad High Risk & 175 & & 76.74±5.49 & 81.92±5.40 & 79.30±4.15 \\
\quad Medium Risk & 297 & & 68.61±3.84 & 92.20±3.31 & 78.98±2.87 \\
\quad Low Risk & 206 & & 89.97±5.34 & 54.93±6.83 & 68.36±5.57 \\
\quad Safe & 8,246 & & 99.95±0.04 & 99.56±0.13 & 99.74±0.07 \\
\addlinespace
\textbf{Figure Consistency} & 554 & 91.2±2.15 & & & \\
\quad Consistent & 371 & & 97.6±1.80 & 86.8±3.65 & 91.9±2.70 \\
\quad Inconsistent & 183 & & 78.1±5.85 & 95.6±3.20 & 86.0±4.35 \\
\bottomrule
\end{tabular}

\begin{tablenotes}
\small
\item Values shown as mean ± 95\% confidence interval half-width from bootstrap resampling (1,000 iterations). 
\end{tablenotes}
\end{table}
\subsubsection{Regulatory Compliance Detection Performance}
The regulatory compliance detection module was evaluated on 10,841 sentences extracted from the aforementioned patents. This dataset exhibits significant class imbalance, with 10,298 compliant sentences (95.0\%) and 543 non-compliant sentences (5.0\%), mirroring the realistic distribution observed in patent prosecution practice.

As shown in Table~\ref{tab:detection_summary}, the module achieved exceptional performance with a balanced accuracy of 99.74±0.07\%. 
Most notably, the perfect recall indicates no regulatory violations are missed, which is particularly valuable when overlooking non-compliant content could have serious legal consequences. The precision of 91.17±2.11\% for non-compliant detection indicates that approximately 8.83\% of flagged content may require manual verification. For compliant content, the model demonstrates perfect precision (100.0±0.00\%) with high recall (99.49±0.14 \%), meaning that content classified as compliant can be trusted with high confidence, while only 0.51\% of actually compliant content is unnecessarily flagged for review.

\textbf{Error Analysis}: The 8.83\% false positive rate in non-compliant detection primarily stems from: (1) borderline cases where regulatory language is technically correct but stylistically unusual, (2) complex conditional statements that the model interprets as potentially non-compliant, and (3) domain-specific terminology variations that deviate from standard regulatory patterns. These errors typically involve nuanced linguistic interpretations and require deep contextual understanding.

\textbf{Practical Implications}: The perfect recall for non-compliant detection ensures comprehensive regulatory coverage, while the moderate false positive rate requires manageable manual review. The module exhibits a tendency to flag more violations, aligning with patent prosecution priorities where undetected violations carry greater consequences. This performance enables automated first-pass screening with human verification for flagged content.

\subsubsection{Technical Coherence Validation Performance}

\begin{table}[htbp]
\centering
\caption{Confusion Matrix Analysis for Risk Category Misclassifications}
\label{tab:confusion_analysis}
\begin{tabular}{lcccc}
\toprule
\textbf{True Label} & \textbf{Predicted as High} & \textbf{Predicted as Medium} & \textbf{Predicted as Low} & \textbf{Predicted as Safe} \\
\midrule
High Risk (175) & \textbf{144 (82.3\%)} & 24 (13.7\%) & 7 (4.0\%) & 0 (0.0\%) \\
Medium Risk (297) & 21 (7.1\%) & \textbf{274 (92.3\%)} & 2 (0.7\%) & 0 (0.0\%) \\
Low Risk (206) & 13 (6.3\%) & 76 (36.9\%) & \textbf{113 (54.9\%)} & 4 (1.9\%) \\
Safe (8246) & 10 (0.1\%) & 23 (0.3\%) & 4 (0.0\%) & \textbf{8209 (99.6\%)} \\
\midrule
\multicolumn{5}{l}{\textbf{Critical Error Analysis}} \\
High→Medium/Low & \multicolumn{4}{l}{31/175 = 17.7\% (High-risk issues under-escalated)} \\
Medium→Low/Safe & \multicolumn{4}{l}{2/297 = 0.7\% (Medium-risk issues under-escalated)} \\
Low→High & \multicolumn{4}{l}{13/206 = 6.3\% (Low-risk issues over-escalated to high)} \\
Safe→High/Medium & \multicolumn{4}{l}{33/8246 = 0.4\% (Safe content over-escalated)} \\
\midrule
\multicolumn{5}{l}{\textbf{Overall Performance Metrics}} \\
Overall Accuracy & \multicolumn{4}{l}{8740/8924 = 97.9\%} \\
Risk Detection Rate & \multicolumn{4}{l}{531/678 = 78.3\% (High+Medium+Low correctly identified)} \\
\bottomrule
\end{tabular}
\begin{tablenotes}
\small
\item Values in parentheses represent the proportion of true labels classified into each predicted category. Diagonal values (in bold) represent correct classifications. 
\end{tablenotes}
\end{table}

The technical coherence validation module was evaluated with 8,924 sentences, derived from the original 10,841 sentences after systematic removal of non-technical template content. The evaluation dataset demonstrates pronounced class imbalance characteristics. The safe content constitutes 8,246 samples (92.4\%), while risk categories represent smaller proportions.

Despite substantial class imbalance, the module achieved a balanced accuracy of 82.12±2.28\%, validating its practical applicability to real patent prosecution scenarios where safe content predominates. Meanwhile, the inherent complexity of technical risk assessment leads to significant performance variations across categories. Specifically, safe content identification achieves near-perfect metrics with F1-score of 99.74±0.07\%, with only 0.4\% requiring unnecessary manual intervention.
Among risk categories, medium-risk content shows the strongest recall at 92.20±3.31\%. High-risk content demonstrates balanced performance with F1-score of 79.30±4.15\%. Low-risk content presents the greatest classification challenge with F1-score of 68.36±5.57\%, as 45.1\% of cases are over-escalated to higher risk categories.

\textbf{Error Analysis}: Several risk level classification challenges require attention. First, high-risk under-escalation accounts for 17.7\% of critical cases. This issue stems  primarily from insufficient recognition of technical terminology modification severity and claim-description mismatch implications. Second, 45.1\% of low-risk issues are over-escalated, primarily involving typographical and formatting problems. Third, the module exhibits risk boundary discrimination difficulties between low and medium risk categories, showing classification bias toward higher risk levels.

\textbf{Practical Implications}: The high accuracy of 99.56\% for safe content enables automated processing of routine content, allowing humans to focus attention on problematic cases. However, the 17.7\% under-escalation rate for high-risk content requires expert validation of critical modifications. The 1.9\% missed risk rate aligns with patent prosecution risk management principles where undetected technical inconsistencies pose greater consequences. Although low-risk over-escalation remains notable, it carries minimal operational impact due to human intervention capabilities.

\subsubsection{Figure-Reference Consistency Verification Performance}

The figure-reference consistency verification module was evaluated on 554 figure-reference pairs extracted from 144 patents, with 371 consistent and 183 inconsistent pairs. With a balanced accuracy of 91.2±2.15\%, the module demonstrated strong capability in figure-text consistency verification tasks. Consistent pairs achieved reliable precision of 97.6±1.80\%. For inconsistent pairs, the module showed excellent detection capability, reaching 95.6±3.20\% recall and 78.1±5.85\% precision.

\textbf{Error Analysis}: Classification errors stem from multimodal processing limitations in complex scenarios. Dense reference numbering (>15) complicates both reference extraction and subsequent visual identification within figures. Meanwhile, non-sequential or nested reference numbering complicates automated matching algorithms. Furthermore, figure quality issues including low resolution or overlapping reference callouts impede accurate element identification.

\textbf{Practical Implications}: The high recall for inconsistency detection (95.6\%) ensures comprehensive identification of genuine figure-reference mismatches, critical for patent quality assurance. While the 21.9\% false positive rate requires additional manual review, this conservative approach prevents overlooking actual inconsistencies. The 91.2\% balanced accuracy supports automated deployment with human verification for complex reference scenarios exceeding 15 elements or involving non-standard numbering patterns.

\textbf{Practical Implications}: The module effectively identifies figure-reference inconsistencies for patent quality assurance with 91.2\% balanced accuracy. High recall (95.6\%) can prevent overlooking actual inconsistencies, while the 21.9\% false positive rate requires manual review of flagged cases.

\subsection{Defect Distribution Analysis (RQ2)}

We analyze defect rate (\%) variations across document sections and IPC classes to identify structural and domain-specific factors influencing patent generation quality.

\subsubsection{Patent Section Analysis}

\begin{table}[htbp]
\centering
\caption{Patent Defect Distribution by Patent Sections}
\label{tab:patent_defect_sections}
\begin{threeparttable}
\begin{tabular}{c|c|c|c|c|c|c}
\hline
\textbf{Section} & \textbf{Non-} & \textbf{High} & \textbf{Medium} & \textbf{Low} & \textbf{Figure} & \textbf{Overall} \\
 & \textbf{Compliance} & \textbf{Risk} & \textbf{Risk} & \textbf{Risk} & \textbf{Inconsist.} & \textbf{Defect} \\
 & \textbf{(\%)} & \textbf{(\%)} & \textbf{(\%)} & \textbf{(\%)} & \textbf{(\%)} & \textbf{(\%)} \\
\hline
Technical Field & \cellcolor{green!30}2.21 & \cellcolor{red!30}15.43 & \cellcolor{green!30}4.94 & \cellcolor{green!30}3.70 & \cellcolor{green!30}0.00 & \cellcolor{red!30}17.51 \\
Background & \cellcolor{green!30}4.02 & \cellcolor{green!30}4.20 & \cellcolor{yellow!30}5.67 & \cellcolor{green!30}1.63 & \cellcolor{green!30}0.00 & \cellcolor{red!30}20.40 \\
Invention Content & \cellcolor{yellow!30}6.22 & \cellcolor{green!30}0.86 & \cellcolor{green!30}3.16 & \cellcolor{green!30}3.66 & \cellcolor{green!30}0.00 & \cellcolor{red!30}21.05 \\
Brief Desc. of Drawings & \cellcolor{red!30}11.40 & \cellcolor{green!30}1.31 & \cellcolor{yellow!30}6.11 & \cellcolor{green!30}3.49 & \cellcolor{red!30}33.03 & \cellcolor{red!30}23.56 \\
Detailed Embodiments & \cellcolor{green!30}4.24 & \cellcolor{green!30}1.44 & \cellcolor{green!30}2.33 & \cellcolor{green!30}1.72 & \cellcolor{green!30}0.00 & \cellcolor{red!30}20.00 \\
\hline
\end{tabular}
\begin{tablenotes}
\small
\item \textbf{Notes:} Color coding: green (<5\%), yellow (5-10\%), red (>10\%).
\end{tablenotes}
\end{threeparttable}
\end{table}

Table \ref{tab:patent_defect_sections} shows that patent quality challenges manifest as distinct error patterns specific to each section's functional requirements.
\textbf{Technical Field} achieves the lowest defect rate at 17.51\%, with robust domain classification accuracy. However, concentrated high-risk defects (15.43\%) are document type misclassification between "invention" and "utility model" categories. 
There are two critical defect patterns in \textbf{Background} sections. First, self-novelty negation occurs when prior art presentation undermines novelty claims. Additionally, low relevance associations arise when referenced literature lacks technical connection to the invention.
With 21.05\% defects identified, the primary issues in \textbf{Invention Content} include technical detail deficiencies and terminological inconsistencies.
\textbf{Brief Description of Drawings} exhibits the highest defect rate at 23.56\%. This concentration occurs because figure-text inconsistency cases, though detected throughout the entire document, are systematically categorized within this section for evaluation efficiency. The resulting high error rate not only reflects our methodological approach but also highlights the need for enhanced multimodal capabilities in current patent drafting tools.
Apart from similar defect patterns in Invention Content, \textbf{Detailed Embodiments} defects also include identical embodiment duplications and misuse of relational terms (e.g., "optional," "further"). These errors can significantly compromise patent validity by failing to demonstrate adequate technical variation and proper hierarchical relationships between embodiments, thereby undermining the enabling disclosure requirement.
Cross-sectional analysis demonstrates that maintaining terminological consistency and figure-text alignment across multiple sections represents a fundamental challenge requiring priority attention in patent drafting.

\subsubsection{Technical Domain Analysis}

\begin{table}[htbp]
\centering
\caption{Patent Defect Distribution by IPC Classification}
\label{tab:patent_quality_ipc}
\begin{threeparttable}
\begin{tabular}{c|c|c|c|c|c|c}
\hline
\textbf{IPC} & \textbf{Non-} & \textbf{High} & \textbf{Medium} & \textbf{Low} & \textbf{Figure} & \textbf{Overall} \\
 & \textbf{Compliance} & \textbf{Risk} & \textbf{Risk} & \textbf{Risk} & \textbf{Inconsist.} & \textbf{Defect} \\
 & \textbf{(\%)} & \textbf{(\%)} & \textbf{(\%)} & \textbf{(\%)} & \textbf{(\%)} & \textbf{(\%)} \\
\hline
A & \cellcolor{yellow!30}6.82 & \cellcolor{green!30}1.83 & \cellcolor{green!30}3.84 & \cellcolor{green!30}2.84 & \cellcolor{red!30}39.66 & \cellcolor{green!30}4.38 \\
B & \cellcolor{yellow!30}5.72 & \cellcolor{green!30}1.87 & \cellcolor{green!30}3.31 & \cellcolor{green!30}2.54 & \cellcolor{red!30}45.83 & \cellcolor{green!30}4.32 \\
C & \cellcolor{yellow!30}5.28 & \cellcolor{green!30}2.58 & \cellcolor{green!30}3.38 & \cellcolor{green!30}3.83 & \cellcolor{red!30}16.67 & \cellcolor{green!30}3.96 \\
D & \cellcolor{green!30}4.85 & \cellcolor{green!30}1.40 & \cellcolor{green!30}3.23 & \cellcolor{green!30}2.26 & \cellcolor{red!30}27.42 & \cellcolor{green!30}3.41 \\
E & \cellcolor{yellow!30}5.83 & \cellcolor{green!30}3.12 & \cellcolor{yellow!30}5.66 & \cellcolor{green!30}3.35 & \cellcolor{red!30}33.82 & \cellcolor{yellow!30}5.09 \\
F & \cellcolor{yellow!30}9.11 & \cellcolor{green!30}3.10 & \cellcolor{green!30}4.94 & \cellcolor{green!30}1.84 & \cellcolor{red!30}32.35 & \cellcolor{yellow!30}5.45 \\
G & \cellcolor{green!30}2.14 & \cellcolor{green!30}1.35 & \cellcolor{green!30}2.21 & \cellcolor{green!30}1.35 & \cellcolor{red!30}32.61 & \cellcolor{green!30}2.12 \\
H & \cellcolor{green!30}4.24 & \cellcolor{green!30}1.31 & \cellcolor{green!30}1.84 & \cellcolor{green!30}1.40 & \cellcolor{red!30}32.50 & \cellcolor{green!30}2.77 \\
\hline
\end{tabular}
\begin{tablenotes}
\small
\item \textbf{Notes:} Color coding: green (<5\%), yellow (5-10\%), red (>10\%)
\end{tablenotes}
\end{threeparttable}
\end{table}

A detailed examination of defect patterns across IPC domains was conducted, revealing significant performance disparities (Table \ref{tab:patent_defect_sections}). Mechanical Engineering requires detailed documentation of three-dimensional systems with intricate component relationships, while Construction involves multi-disciplinary integration across structural, mechanical, electrical, and environmental subsystems. These comprehensive requirements make claim-specification alignment and figure-text consistency particularly challenging.
Physics (Class G) achieves consistently superior performance with the lowest overall defect rate at 2.12\%, leading in non-compliance (2.14\%), medium-risk (2.21\%), and low-risk (1.35\%) categories.
This exceptional performance can be attributed to mathematical precision, standardized measurement protocols, and well-defined theoretical foundations that create natural alignment between theoretical descriptions and practical implementations. Moreover, Physics patents typically focus on method-based inventions rather than complex apparatus descriptions, thereby reducing the reference numeral management complexity that challenges other technical domains.
Across all IPC classes, figure inconsistency rates range from 16.67\% to 45.83\%, representing the most pervasive quality challenge in patent documentation.


\subsection{Comparative Quality Analysis (RQ3)}

Average defects per document is used to compare patent quality across sources to capture real-world user experience despite document length variations. Table \ref{tab:quality_comparison} compares human-authored and AI-generated patents with statistical significance testing for all pairwise comparisons.

\begin{table}[h]
\centering
\begin{threeparttable}
\caption{AI Tools vs Human Performance in Patent Drafting Quality}
\label{tab:quality_comparison}
\begin{tabular}{lcccccc}
\toprule
\textbf{Dimension} & \textbf{Human} & \textbf{Eureka} & \textbf{MindFlowing} & \textbf{H-E} & \textbf{H-M} & \textbf{E-M} \\
\midrule
\textbf{Regulatory Compliance} & & & & & & \\
\quad Non-compliance & 2.700 & 4.400 & 3.775 & -1.700$^a$ & -1.075$^a$ & +0.625$^a$ \\
\addlinespace
\textbf{Technical Coherence} & & & & & & \\
\quad High Risk & 0.325 & 1.900 & 1.825 & -1.575$^a$ & -1.500$^a$ & +0.075$^a$ \\
\quad Medium Risk & 2.050 & 1.050 & 2.275 & +1.000$^a$ & -0.225$^a$ & -1.225$^a$ \\
\quad Low Risk & 1.913 & 0.475 & 0.850 & +1.438$^a$ & +1.062 & -0.375$^a$ \\
\addlinespace
\textbf{Figure-Text Consistency} & & & & & & \\
\quad Figure-Ref Inconsistency  & 0.139 & 1.212 & 3.410 & -1.073$^b$ & -3.271$^b$ & -2.198$^b$ \\
\midrule
\textbf{Total Defects per Doc} & \textbf{7.126} & \textbf{9.037} & \textbf{12.135} & \textbf{-1.911}$^b$ & \textbf{-5.009}$^b$ & \textbf{-3.098}$^b$ \\
\bottomrule
\end{tabular}
\begin{tablenotes}[flushleft]
\small
\item \textbf{Notes:} $^a$ p<0.001, $^b$ p<0.05
\end{tablenotes}
\end{threeparttable}
\end{table}

The human-AI gaps show statistical significance (H-E: -1.911, H-M: -5.009, p<0.05),  indicating patent drafting tools require ongoing enhancement.
Meanwhile, quality assurance remains crucial regardless of the drafting method, given the presence of defects in all patent types.
Figure-reference consistency exhibits the largest relative quality gaps (H-E: -1.073, H-M: -3.271, p<0.05) and represents the most challenging aspect for AI patent drafting tools. Furthermore, the substantial difference between Eureka and MindFlowing (E-M: -2.198, p<0.05) highlights the necessity of multi-modal LLMs in patent drafting, as MindFlowing's figure-free generation leads to misaligned references while Eureka achieves better figure-text coordination. Yet substantial enhancement remains needed to reach human-level accuracy.
In terms of technical coherence analysis, human-authored patents show notably more low-risk defects, primarily typographical mistakes and minor formatting issues that do not compromise technical understanding. More critically, both patent drafting tools exhibit significantly more defects in higher-risk categories (Human: 0.325, Eureka: 1.900, MindFlowing: 1.825). These defects primarily manifest as inconsistencies between patent sections and misalignment with claims, revealing fundamental architectural limitations.
All sources show substantial regulatory compliance challenges, though the defect patterns differ significantly. Human-authored patents primarily contain typographical errors, punctuation mistakes, and minor commercial language issues that escaped editorial review. 
In contrast, AI-generated compliance issues predominantly involve inappropriate commercial language usage, claims of beneficial effects without adequate technical support, and inconsistent patent classification terminologies that create regulatory ambiguity. 


\section{Conclusion}
\label{sec:conclusion}

We present an automated patent quality assessment framework to address the limitations of manual patent review and emerging quality challenges brought by AI-assisted drafting. Our framework comprises three detection modules for regulatory compliance, technical coherence, and figure-reference consistency, and one integration module offering improvement guidance. To validate its effectiveness, we constructed a comprehensive dataset containing 80 human-authored and 80 AI-generated patents from two patent drafting tools. Experimental results show that the detection modules achieve balanced accuracies of 99.74\%, 82.12\%, and 91.2\%, respectively, when evaluated against expert annotations. Additionally, we investigate the distribution of defects across patent sections, technical domains, and authoring sources. Our findings demonstrate that patent drafting tools have not yet achieved human-level quality, with significant gaps in figure-text and claim-specification alignment in domains involving complex multi-component architectures.
Several limitations should be acknowledged. First, the dataset size of 160 patents, while comprehensive, remains limited for broader generalization. Second, the technical coherence module cannot distinguish between severity levels accurately, and the figure-text consistency detection module requires further refinement. 
This research can facilitate manual patent examination while providing enhancement directions for AI-generated patents. Additionally, the multimodal and claim-specification alignment challenges identified above can be research priorities for future AI-assisted patent drafting.

\bibliographystyle{unsrt}  
\bibliography{main}

\begin{thebibliography}{10}

\bibitem{rehnstrom2024drafting}
Dante Rehnstr{\"o}m and Alex Johansson.
\newblock Drafting the future: Genai in patents.
\newblock 2024.

\bibitem{mehrotra2024navigating}
Ananya Mehrotra.
\newblock Navigating the intellectual property landscape in the age of artificial intelligence: Towards a global legal paradigm.
\newblock {\em International Journal of Innovations in Science, Engineering And Management}, pages 253--258, 2024.

\bibitem{bui2025advancing}
Luong~Vu Bui.
\newblock Advancing patent law with generative ai: Human-in-the-loop systems for ai-assisted drafting, prior art search, and multimodal ip protection.
\newblock {\em World Patent Information}, 80:102341, 2025.

\bibitem{khera2025can}
Bhakti Khera, Rezvan Alamian, Pascal~A Scherz, and Stephan~M Goetz.
\newblock Can large language models understand as well as apply patent regulations to pass a hands-on patent attorney test?
\newblock {\em arXiv preprint arXiv:2507.10576}, 2025.

\bibitem{ray2023ai}
Ananya Ray.
\newblock Ai in ipr: Leveraging technology for efficiency and addressing concerns.
\newblock {\em Ananya Ray, AI in IPR: Leveraging Technology for Efficiency and Addressing Concerns, Ile Lex Speculum (Ile Ls)}, 1(1):333--341, 2023.

\bibitem{wang2024patentformer}
Juanyan Wang, Sai Krishna~Reddy Mudhiganti, and Manali Sharma.
\newblock Patentformer: a novel method to automate the generation of patent applications.
\newblock In {\em Proceedings of the 2024 conference on empirical methods in natural language processing: industry track}, pages 1361--1380, 2024.

\bibitem{ren2025large}
Runtao Ren, Jian Ma, and Jianxi Luo.
\newblock Large language model for patent concept generation.
\newblock {\em Advanced Engineering Informatics}, 65:103301, 2025.

\bibitem{burke2007measuring}
Paul~F Burke and Markus Reitzig.
\newblock Measuring patent assessment quality—analyzing the degree and kind of (in) consistency in patent offices’ decision making.
\newblock {\em Research Policy}, 36(9):1404--1430, 2007.

\bibitem{harhoff2003citations}
Dietmar Harhoff, Frederic~M Scherer, and Katrin Vopel.
\newblock Citations, family size, opposition and the value of patent rights.
\newblock {\em Research policy}, 32(8):1343--1363, 2003.

\bibitem{hu2023evaluation}
Zewen Hu, Xiji Zhou, and Angela Lin.
\newblock Evaluation and identification of potential high-value patents in the field of integrated circuits using a multidimensional patent indicators pre-screening strategy and machine learning approaches.
\newblock {\em Journal of Informetrics}, 17(2):101406, 2023.

\bibitem{trappey2019patent}
Amy~JC Trappey, Charles~V Trappey, Usharani~Hareesh Govindarajan, and John~JH Sun.
\newblock Patent value analysis using deep learning models—the case of iot technology mining for the manufacturing industry.
\newblock {\em IEEE Transactions on Engineering Management}, 68(5):1334--1346, 2019.

\bibitem{zhang2024research}
Liwei Zhang, Tongtong Zhang, Yutao Lang, Jiaxi Li, and Fujun Ji.
\newblock Research on patent quality evaluation based on rough set and cloud model.
\newblock {\em Expert Systems with Applications}, 235:121057, 2024.

\bibitem{wang2024autopatent}
Qiyao Wang, Shiwen Ni, Huaren Liu, Shule Lu, Guhong Chen, Xi~Feng, Chi Wei, Qiang Qu, Hamid Alinejad-Rokny, Yuan Lin, et~al.
\newblock Autopatent: a multi-agent framework for automatic patent generation.
\newblock {\em arXiv preprint arXiv:2412.09796}, 2024.

\bibitem{zhao2023survey}
Wayne~Xin Zhao, Kun Zhou, Junyi Li, Tianyi Tang, Xiaolei Wang, Yupeng Hou, Yingqian Min, Beichen Zhang, Junjie Zhang, Zican Dong, et~al.
\newblock A survey of large language models.
\newblock {\em arXiv preprint arXiv:2303.18223}, 1(2), 2023.

\bibitem{yoo2025patentscore}
Yongmin Yoo, Qiongkai Xu, and Longbing Cao.
\newblock Patentscore: Multi-dimensional evaluation of llm-generated patent claims.
\newblock {\em arXiv preprint arXiv:2505.19345}, 2025.

\bibitem{jiang2024can}
Lekang Jiang, Caiqi Zhang, Pascal~A Scherz, and Stephan Goetz.
\newblock Can large language models generate high-quality patent claims?
\newblock {\em arXiv preprint arXiv:2406.19465}, 2024.

\bibitem{lee2020patent}
Jieh-Sheng Lee and Jieh Hsiang.
\newblock Patent claim generation by fine-tuning openai gpt-2.
\newblock {\em World Patent Information}, 62:101983, 2020.

\bibitem{jiang2024patent}
Lekang Jiang, Pascal~A Scherz, and Stephan Goetz.
\newblock Patent-cr: A dataset for patent claim revision.
\newblock {\em arXiv preprint arXiv:2412.02549}, 2024.

\bibitem{grimaldi2020indexes}
Michele Grimaldi and Livio Cricelli.
\newblock Indexes of patent value: a systematic literature review and classification.
\newblock {\em Knowledge Management Research \& Practice}, 18(2):214--233, 2020.

\bibitem{cricelli2021patent}
Livio Cricelli, Michele Grimaldi, Francesco Rogo, and Serena Strazzullo.
\newblock Patent ranking indicators: a framework for the evaluation of a patent portfolio.
\newblock {\em International Journal of Intellectual Property Management}, 11(2):185--218, 2021.

\bibitem{wu2016patent}
Jheng-Long Wu, Pei-Chann Chang, Cheng-Chin Tsao, and Chin-Yuan Fan.
\newblock A patent quality analysis and classification system using self-organizing maps with support vector machine.
\newblock {\em Applied soft computing}, 41:305--316, 2016.

\bibitem{jiang2025towards}
Lekang Jiang, Pascal~A Scherz, and Stephan Goetz.
\newblock Towards better evaluation for generated patent claims.
\newblock {\em arXiv preprint arXiv:2505.11095}, 2025.

\bibitem{gerken2012new}
Jan~M Gerken and Martin~G Moehrle.
\newblock A new instrument for technology monitoring: novelty in patents measured by semantic patent analysis.
\newblock {\em Scientometrics}, 91(3):645--670, 2012.

\bibitem{jang2023explainable}
Hyejin Jang, Sunhye Kim, and Byungun Yoon.
\newblock An explainable ai (xai) model for text-based patent novelty analysis.
\newblock {\em Expert systems with applications}, 231:120839, 2023.

\bibitem{du2021personalized}
Wei Du, Yibo Wang, Wei Xu, and Jian Ma.
\newblock A personalized recommendation system for high-quality patent trading by leveraging hybrid patent analysis.
\newblock {\em Scientometrics}, 126(12):9369--9391, 2021.

\bibitem{wu2024surveyllmgeneratedtextdetection}
Junchao Wu, Shu Yang, Runzhe Zhan, Yulin Yuan, Derek~F. Wong, and Lidia~S. Chao.
\newblock A survey on llm-generated text detection: Necessity, methods, and future directions, 2024.

\bibitem{arase2013machine}
Yuki Arase and Ming Zhou.
\newblock Machine translation detection from monolingual web-text.
\newblock In {\em Proceedings of the 51st Annual Meeting of the Association for Computational Linguistics (Volume 1: Long Papers)}, pages 1597--1607, 2013.

\bibitem{wu2024detectrl}
Junchao Wu, Runzhe Zhan, Derek~F Wong, Shu Yang, Xinyi Yang, Yulin Yuan, and Lidia~S Chao.
\newblock Detectrl: Benchmarking llm-generated text detection in real-world scenarios.
\newblock {\em arXiv preprint arXiv:2410.23746}, 2024.

\bibitem{mitchell2023detectgpt}
Eric Mitchell, Yoonho Lee, Alexander Khazatsky, Christopher~D. Manning, and Chelsea Finn.
\newblock Detectgpt: Zero-shot machine-generated text detection using probability curvature, 2023.

\bibitem{kirchenbauer2023watermark}
John Kirchenbauer, Jonas Geiping, Yuxin Wen, Jonathan Katz, Ian Miers, and Tom Goldstein.
\newblock A watermark for large language models.
\newblock In {\em International Conference on Machine Learning}, pages 17061--17084. PMLR, 2023.

\end{thebibliography}

\end{document}